# MUDANÇAS CLIMÁTICAS E CIÊNCIAS SOCIAIS: UMA ANÁLISE BIBLIOMÉTRICA CLIMATE CHANGE AND SOCIAL SCIENCES: A BIBLIOMETRIC ANALYSIS

FLÁVIO DIAS DE MORAES, ANA LIA LEONEL, PEDRO HENRIQUE TORRES, PEDRO ROBERTO JACOBI, SANDRA MOMM

**Flávio Campopiano Dias de Moraes** is a physicist and Ph.D. in Physics, currently conducting Postdoctoral Research in Integrated Photonic Devices at the Gleb Wataghin Physics Institute, of the State University of Campinas - UNICAMP, Brazil, where he studies artificial intelligence and neural networks. fmoraes@if.usp.br

**Ana Lia Leonel** has a bachelor's degree in Social Sciences and a Master's in Territory Planning and Management. She is a researcher in the Graduate Program in Territory Planning and Management at the Federal University of ABC, São Bernardo do Campo, Brazil, developing research in the areas of environmental planning, territorial planning, metropolitan governance, public policies and environmental sociology regarding climate change. analia@ufabc.edu.br

**Pedro Henrique Campello Torres** is a social scientist and Ph.D. in Social Sciences and conducts Postdoctoral research in Environmental Science at the Institute of Energy and Environment at the University of São Paulo, Brazil. His research topics are urban planning and the environment, environmental sociology, urban sociology, climate change, and environmental justice. pedrotorres@usp.br

**Pedro Roberto Jacobi** is a social scientist and Ph.D. in Sociology. He is a Senior Professor of the Graduate Program in Environmental Science at the Institute of Energy and Environment of the University of Sao Paulo, Brazil. He conducts research on environmental governance, education for sustainability, and climate change. prjacobi@gmail.com

**Sandra Momm** is an architect and urbanist and has a Ph.D. in Environmental Science. She is an Adjunct Professor at the Federal University of ABC where she coordinates the Postgraduate Program in Territory Planning and Management. She conducts research on territorial planning in interface with environmental issues such as climate change, water resources, and protected areas. sandra.momm@ufabc.edu.br

**Abstract**

The complexity of emergent wicked problems, such as climate change, culminates in a reformulation of how we think about society and mobilize scientists from various disciplines to seek solutions and perspectives on the problem. From an epistemological point of view, it is essential to evaluate how such topics can be developed inside the academic arena but, to do that, it is necessary to perform complex analysis of the great number of recent academic publications. In this work, we discuss how climate change has been addressed by social sciences in practice. Can we observe the development of a new epistemology by the emergence of the climate change debate? Are there contributions in academic journals within the field of social sciences addressing climate change? Which journals are these? Who are the authors? To answer these questions, we developed an innovative method combining different tools to search, filter, and analyze the impact of the academic production related to climate change in social sciences in the most relevant journals.

**Keywords:** Bibliometric analysis, Algorithms, Climate change, Social Sciences, Natural Sciences

## 1 Introduction

There has been an increasingly widespread idea that social sciences should be considered in processes notoriously marked by the production of specific knowledge from natural sciences. This agenda has become more relevant in academia, public policies, international forums, and agencies of the United Nations and their scientific reports, such as the Intergovernmental Panel on Climate Change (IPCC, 1990, 1995, 2001, 2007, 2013).

On July 4th, 2018, a historic meeting took place in Paris (France) leading to the merger of the International Council for Science (ICSU) and the International Social Science Council (ISSC). It launched the International Science Council (ISC), an interdisciplinary forum that aims to bring together experts from all continents, looking for an integration between natural and social sciences. At the ISC launch, the president of the former ICSU and Secretary of the French Academy of Sciences, Catherine Brechignac, stated that "natural sciences should no longer dictate the research agenda of the earth system sciences, the social sciences must have at least the same role of the natural sciences[1].

It is no longer a matter of discussing whether the contributions from the social sciences field are important to address climate change but how is their construction, reception, and circulation in the academic arena, especially within the spectrum of our field. In this sense, the search for a clear and imperative dialogue for the construction of new “hybrid understandings” (Jacobi, Rotondaro and Torres, 2019, our translation) in a contemporary “world in metamorphosis” (Beck, 2018, our translation) is a contribution from the social sciences field. How is this happening in practice? Is the production of knowledge in the Social Sciences field moving along with these demands? To identify these changes, we propose a profound analysis of the current scientific production related to climate change published in the most relevant journals with studies in social sciences.

The initial idea of this work was to combine social sciences, environmental sciences, and territorial planning to perform an interdisciplinary analysis of the production on climate change. The difficulty of finding appropriate tools to filter and analyze the most relevant productions for the research led to the necessity of combining computer science and data analyses to develop a new method for conducting these processes. Since the lack of tools itself is a truly relevant subject that limits the research process, the developed method became the central focus of this work. This article describes a set of tools and procedures for a new method specifically developed for an analysis focused on organizing the knowledge produced by the social science community within the interdisciplinary dialogue about climate change.

## 2 Materials and Methods

Considering that the subject under discussion is recent and has not clearly defined theoretical lines or academic tradition, the bibliographic research method (State of the art review; Citation Analysis and others) was considered the most appropriate (Ferreira, 2002; Matsuoka and Kaplan, 2008; Creswell, 2010; Lecy and Beatty, 2012; Sanchez, 2017). The analysis of the current scientific production related to climate change and social science in high impact journals demands determining the most relevant journals that are publishing in the social sciences. The absence of works related to climate change in such journals is also a relevant data.

The Journal Impact Factor (JIF), which is published annually in the Journal Citation Reports (JCR), was used as a parameter to determine the journal's relevance. However, due to interdisciplinarity, the evaluation of the journal's main subject is not enough to determine if it frequently publishes studies related to social sciences. If we filter a survey by the journal subject area, we could undesirably neglect important papers published as exceptions in journals from different areas. The Clarivate Analytics search base, Web of Science, does not consider the area of individual documents but only the journal's main subject area. Therefore, we used the Scopus search database to verify how many studies of social science were published by each journal in a specific time window. Scopus considers, for each journal, how many articles are indexed in the subject areas, and there are cases in which one article can belong to more than one subject area.

The combined information of the JCR and Scopus database is enough to rank the most relevant journals that frequently publish social science studies. However, the obtainment of this kind of information can be tricky as it requires the search of each journal on JIF ranking through the Scopus database. The data compilation for the new ranking was retrieved by using a computer bot developed to automatically perform hundreds of searches in the Scopus database. The first journals from the reduced ranking were individually analyzed with VOSviewer to verify the importance of climate change in comparison with other subjects.

A more detailed explanation of the search bot and the VOSviewer analysis is described further in this section. We initially used the method presented in this work to analyze data from 2006 to 2018. The year 2006 was chosen as a chronological framework to include works from a year before the publication of the fourth report of the Intergovernmental Panel on Climate Change (IPCC), which had a great impact on subsequent publications. We include an update of the results in the discussion session.

## 2.1 Data Filtering

The data filtering method consists of using a bot to filter the journals from JCR ranking which, according to Scopus, have published studies indexed in the Social Science subject area during the selected time window. The rank was directly downloaded from the Clarivate Analytics webpage[2] in CSV format. The bot was developed in Python to search in Scopus for every journal from the rank as source title, and to verify if there was at least one document in the Social Science subject area.

Since the Scopus webpage uses JavaScript to render the required information, it was necessary to use automated test software to control the web browser. We used the ChromeDriver software, controlled by the selenium web driver. The unrestricted access to Scopus was guaranteed by the VPN network of the University of São Paulo (USP). The result of the program was compiled in a file with a CSV list containing the journal's name, its position in the JCR rank, the JIF, and the search result tag, which we explain further.

The bot searched on Scopus by directly filling the search parameters (periodic name and date restrictions) in the Scopus search URL. However, it needed to take some care to ensure the reliability of the results: since we are dealing with automatized processes, we have to be sure that all the documents Scopus may find are from the correct journal. To avoid data contamination from other journals with similar names, the search must be limited to the exact source title. This procedure, however, increases the probability of not finding a journal. A series of more complex procedures were necessary to repeat the search of journals that were not found in the first attempt and increase search success without losing reliability. Figure 1 illustrates the main steps and decision flow of the whole search procedure, which are:

**1 Retrieving the journal name from the JIF rank:** we took the journal name from a CSV file exported from the 2017 JCR and accessed it via pandas library;

**2 Adjusting the journal name:** Scopus does not accept non-alphanumeric characters on journal names. It uses 'and' instead of '&' and empty spaces instead of hyphens and slashes;

**3 Creating an URL for the exact source name search:** in the first search process, the bot defined the search as the exact source title by adding *s=EXACTSRCTITLE(Journal+Name)*. It restricted the time through the term*+AND+PUBYEAR+>+2005*. It was also necessary to limit results to the exact source title to ensure there was no data contamination, which was done by the addition of the term*cluster=scoexactsrctitle,"Journal+Name",t*. It is important to consider, for this last procedure, that Scopus search is case sensitive, which introduces some difficulties to the search because JCR rank does not follow any rule to discriminate the use of capital or small letters. The use of Python *.title()* string method may solve this problem in most cases, as in "Energy & Environmental Science", with the URL: *results.uri?*

*src=s&sot=a&s=EXACTSRCTITLE(energy+and+environmental+science)+AND+PUBYEAR+>+2005&cluster=scoexactsrctitle,"Energy+And+Environmental+Science",t.*

However, this solution fails for cases as the "JAMA-JOURNAL OF THE AMERICAN MEDICAL ASSOCIATION", in which the name JAMA should be kept with all uppercase letters. The impossibility of programming solutions for each case makes the program unable to find some of the journals during this first search process. This problem can be fixed by some search special procedures;;

**4 Checking for subject area indexing:** when the search is successful, it is necessary to verify if there are any documents indexed in the Social Science subject area. The bot does this by searching, in the HTML code, for all span elements with class equals to *"btnText"*, which are child from the HTML unordered list element with id equals to *"cluster_SUBJAREA"*. If no document is indexed in Social Science, the journal can be dismissed;

**5 Special procedures:** the special procedures are applied only for journals that were not found during the first search process. The bot carried out a second search without limiting results to the exact source title and then extracted the subject area of the results. If the search does not find any document, the journal is tagged as "not found" for further manual check. If it finds, it is necessary to consider that there may be data contamination. The program then checks if there are documents indexed in Social Science. If there are not, the journal is dismissed despite the possibility of data contamination. However, if there are documents of Social Science, additional procedures are applied:

**a Checking for source title:** it verifies whether the documents are all from the same journal or not. In case of a positive result, the journal does contain documents indexed in Social Science. If negative, we have to look for different source titles.

**b Comparing journal names with source title list:** in case of multiple source titles, the bot compares each title with the journal name we want to check. If the intersection between the words from the source title and the words from the journal name has more than 75% of the words for both names, it is considered a match (even though the intersection between *Nature Materials* and *Nature* contains 100% of the words from Nature, it contains only 50% from *Nature Materials*, thus it is not considered as a match). The comparison was conducted by using small letters only to avoid case-sensitivity. In case of matches, the bot performs a third search by using the source title exactly as it is in Scopus source title list. This search always gives a result which can contain or not documents indexed in Social Science. Depending on the case, the journal was tagged as "Probably OK" and "Probably False" and the name of the searched journal is added to the output data. In case of mismatch, the journal was tagged as "Unsure" for further manual check.

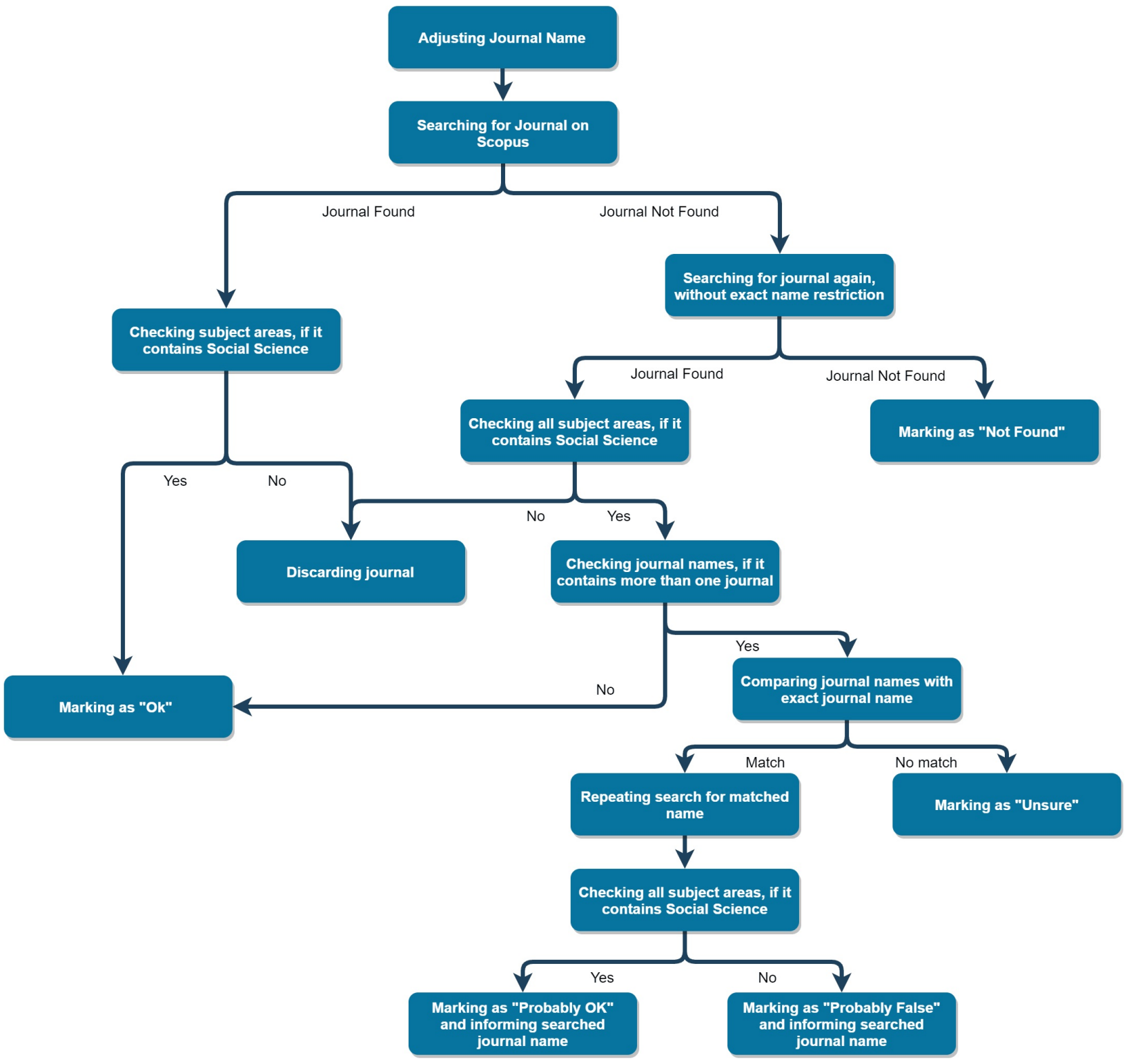

**Fig. 1:** Flowchart diagram representation of the main operation processes of the bot search algorithm for a journal name. Source: The authors, 2020.

## 2.2 Post-filtering and analysis method

The analysis of the relevance of the Social Science subject for the filtered journal began by collecting information on how many papers from the journal were indexed in the Social Science subject area. We did this through a Scopus analysis of all documents from each filtered journal. The journals with less than 1% of documents indexed in Social Science were discarded from further analysis because they cannot be considered as a metric for the social science research community. One percent may seem a small number but it represents a high absolute number of documents when considering the whole number of publications from a journal over a large time.

The result list was limited to the 15 journals with the highest impact factor. Then, the number of documents, published in each of these journals, indexed in Social Science was compared with the number of documents indexed to all other areas. It consequently created a relative index regarding the importance of the journal for the social science scientific community and the influence of the social science community to the journal.

Scopus was once again used to analyze how many articles of each journal from the resulting list contained Climate Change as a keyword. This information, along with the previously discussed information retrieved from the journals, were compiled in Table 1. Then, the VOSviewer[3]application was applied to analyze the keywords of journals with more than 10 documents indexed to Social Science and which contained Climate Change as a keyword. The VOSviewer generated a word cloud from a list of words. In this cloud, the words with higher occurrence are represented with a larger size and the related words appear in clusters. This organization allows us to analyze the relevance of Climate Change as a subject for these journals, which keywords are related to Climate Change, and also which are the other relevant keywords to journals not related to Climate Change.

In a final step, we used Scopus to analyze documents published by the 15 selected journals, between 2007 and 2018, which were indexed in Social Science and contained Climate Change as a keyword. We discuss the result of this analysis in the following section.

# 3 Results and discussion: the relevance and connections of climate change in top-ranked journals

The filtering mechanism was set to stop after tagging 50 journals in the output list, which happened after the bot searched for 804 journals from the JCR rank. From these 50 journals, 31 were correctly found, 17 were tagged as not found, and 2 were tagged with unsure results. It may seem that manually searching for 17 journals is hard work, but it is feasible in contrast with searching through more than 800 journals. The results of the procedures described in the material and methods section are compiled in Table 1. The table describes the 15 journals with the highest impact factors containing more than 1% of papers indexed to Social Science.

| Rank | Journal Title | JIF | Papers Number | SS Papers Number | SS Relative Index | Occurrence of "Climate Change" |
|---|---|---|---|---|---|---|
| **78** | Nature Climate Change | 19.181 | 2192 | 2192 | 50 % | 676 |
| **116** | Behavioral And Brain Sciences | 15.071 | 2688 | 952 | 9,5% | 0 |
| **167** | MMWR-Morbidity And Mortality Weekly Report | 12.888 | 434 | 318 | 22,9% | 0 |
| **232** | Dialogues In Human Geography | 10.214 | 360 | 360 | 100,0% | 4 |
| **339** | Review Of Educational Research | 8.241 | 300 | 300 | 100,0% | 0 |
| **411** | Land Degradation & Development | 7.270 | 1331 | 1317 | 33,1% | 108 |
| **454** | Progress In Human Geography | 6.885 | 730 | 730 | 100,0% | 20 |
| **460** | Journal Of Service Research | 6.842 | 358 | 356 | 33,3% | 0 |
| **467** | Annual Review Of Sociology | 6.773 | 305 | 305 | 100,0% | 0 |
| **525** | Economic Geography | 6.438 | 242 | 242 | 50,0% | 3 |
| **535** | Global Environmental Change-Human And Policy Dimensions | 6.371 | 1425 | 1307 | 47,8% | 632 |
| **570** | Social Issues And Policy Review | 6.143 | 95 | 95 | 50,0% | Null |
| **609** | ISPRS Journal Of Photogrammetry And Remote Sensing | 5.994 | 1588 | 275 | 4,2% | 41 |
| **622** | Tourism Management | 5.921 | 1998 | 1998 | 50,0% | 29 |
| **628** | Administrative Science Quarterly | 5.878 | 285 | 283 | 49,9% | 0 |

**Table 1:** Most relevant journals in social science. JIF: Journal Impact Factor | Paper Number: the total number of papers in the journal from 2006 to current days | SS Paper Number: the total number of papers indexed in the social sciences subject area of the journal from 2006 to current days | SS Relative Index: number of social sciences indexations in relation to other subject areas | Occurrence of "Climate Change": number of times the keyword "Climate Change" appears in the journal from 2006 to 2018 (inclusive) | Null: non-available information. Source: The authors, 2018.

## 3.1 Number of publications indexed as from social science

The possibility of indexing articles on different subject areas, regardless of the journal's main subject area, not only helps researchers to better find and filter bibliography in a systematic review but also brings valuable information about the development of a specific area. This question becomes even more important due to interdisciplinarity. Brint points out the interest of American universities in following "new directions" by seeking "interdisciplinary creativity" (Brint, 2005). According to him, while public universities adopt interdisciplinary strategies "particularly drawn to projects that serve the economic development of their respective States", private universities "are more likely to emphasize the intellectual excitement associated with new fields" (Brint, 2005, pp. 29). The efforts to promote interdisciplinarity had many impacts over scientific production and the "diffusion of research across disciplines in the humanities and social sciences" (Jacobs and Frickel, 2009, pp. 43). The consequences directly affect the journal indicators based on citations: authors that work on interdisciplinary problems tend to publish less papers but receive more citations (Leahey, Beckman and Stanko, 2017). Interdisciplinary problems are challenging and attract the interest of many researchers from different areas. At the same time, the most influential journals, already well-established in their fields, may lose influence over their areas and start undesirable competitions by focusing on interdisciplinary studies. Thus, they prefer to limit the indexation of their publications to their fields.

Table 1 shows that some journals index all papers in the same subject area. The documents from *Nature Climate Change*, for example, are not individually indexed. Instead, they are all indexed in Social Science and Environmental Science. Other magazines, like *Progress in Human Geography* and *Annual Review of Sociology*, index all publications in Social Science only. However, documents from the *Behavioral and Brain Sciences* and the *ISPRS Journal of Photogrammetry and Remote Sensing* would not have been considered if they were not individually indexed since none of those journals are from Social Science. Both journals present a relatively high number of publications on Social Science. The ISPRS also has many of these publications discussing climate change. However, we obtained these results in 2018. If someone tries to repeat this survey now, the result would not contain these two journals since both changed the indexing strategy to the same one adopted by the most influential journals. Table 2 shows the updated data for the same 15 journals. The most significant alterations beyond the aforementioned are the change on JIF and, consequently, the ranking position of some journals.

| Rank | Journal Title | JIF | Papers Number | SS Papers Number | SS Relative Index | Occurrence of "Climate Change" |
|---|---|---|---|---|---|---|
| 65 | Nature Climate Change | 21.722 | 2,537 | 2,537 | 50% | 739 |
| 102 | Behavioral And Brain Sciences | 17.194 | 3426 | 0 | 0% | 7 |
| 138 | MMWR-Morbidity And Mortality Weekly Report | 14.874 | 4186 | 4186 | 25% | 0 |
| 1723 | Dialogues In Human Geography | 3.875 | 417 | 417 | 100% | 11 |
| 327 | Review Of Educational Research | 8.985 | 330 | 330 | 100% | 0 |
| 1406 | Land Degradation & Development | 4.275 | 1623 | 1623 | 33,3% | 144 |
| 579 | Progress In Human Geography | 6.576 | 826 | 826 | 100% | 24 |
| 1562 | Journal Of Service Research | 4.071 | 407 | 407 | 33,3% | 0 |
| 1040 | Annual Review Of Sociology | 4.915 | 335 | 335 | 100% | 1 |
| 537 | Economic Geography | 6.861 | 262 | 262 | 50% | 3 |
| 259 | Global Environmental Change-Human And Policy Dimensions | 10.427 | 1453 | 1453 | 50% | 674 |
| 339 | Social Issues And Policy Review | 8.733 | 105 | 105 | 50% | Null |
| 519 | ISPRS Journal Of Photogrammetry And Remote Sensing | 6.942 | 1846 | 0 | 0% | 54 |
| 707 | Tourism Management | 6.012 | 2232 | 2232 | 50% | 38 |
| 393 | Administrative Science Quarterly | 8.024 | 317 | 317 | 50% | 0 |

**Table 2:** Information update about the journals in Table 1. Source: The authors, 2020.

### 3.2 Occurrence of the "Climate Change" keyword

To answer how Social Sciences are dealing with the Climate Change issue, the first procedure adopted was the analysis of the relevance and connections of Climate Change with the other keywords in the journals from Table 1. *Dialogues in Human Geography* journal indexed all its papers in Social Sciences only. It has also published only 4 Climate Change papers since 2006, which have 16 links to other keywords. Its most common keywords and their frequency are Theoretical Study (62), Human Geography (55), and Economic Geography (28). The *Journal of Photogrammetry and Remote Sensing* keywords are mostly about Remote Sensing (615), Algorithm (390), Accuracy Assessment (356), and Data Set (350). The Social Sciences area has less significance. The 41 papers about Climate Change are indexed in Computer Science, Earth and Planetary Sciences, Engineering, and Physics and Astronomy, and only 9 are also indexed as Social Sciences. These 41 papers were produced mostly in the United States (14) and China (11).

Five journals of Table 1 were selected for containing, at least, 10 papers indexed in Social Science with Climate Change as a keyword. The word clouds below (Figures 2-6) illustrate the keywords from these journals from 2006 to 2018. They were produced with VOSviewer software by selecting all keywords with a minimum of 5 occurrences. Their color scale represents the evolution of the terms in time.

I. Nature Climate Change Journal

*Nature Climate Change* is the highest-ranked journal and the one with most occurrences of Climate Change as a keyword. From 2381 keywords with a minimum of five occurrences, Climate Change is the first on the list, with 676 occurrences and 491 links with other keywords. All papers about Climate Change are indexed in Social Sciences and Natural Sciences. Figure 2 shows the relation among all keywords indexed in the *Nature Climate Change* journal from 2006 to 2018.

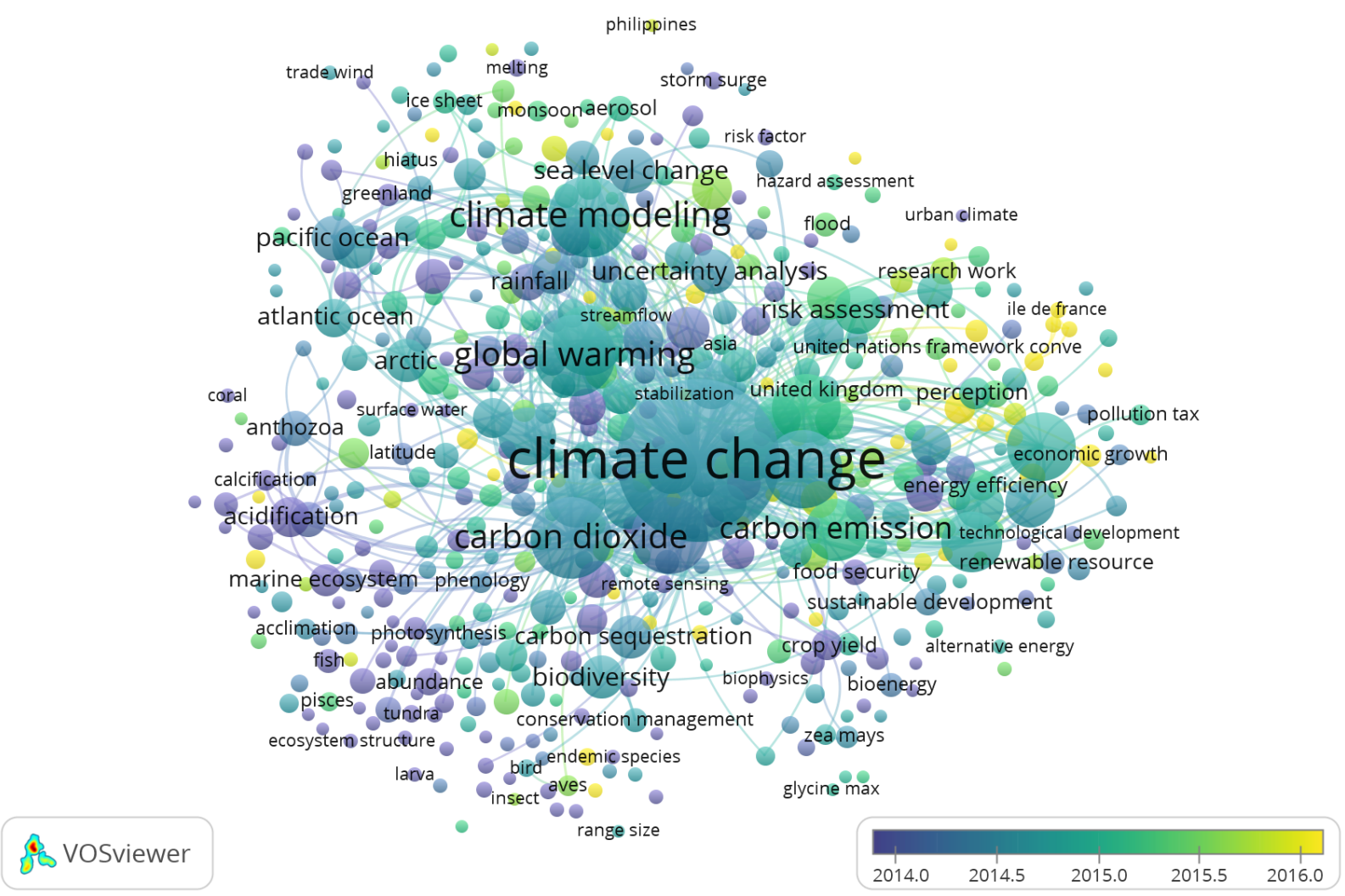

**Fig. 2:** Word cloud of keywords from *Nature Climate Change* journal papers from 2006 to 2018. Source: The authors, 2018.

II. Land Degradation & Development

The *Land Degradation & Development* journal has 108 occurrences of Climate Change as a keyword. This word cloud (Figure 3) has 231 keywords and Climate Change has 190 links. According to Scopus database, all these 108 papers are indexed simultaneously in Agricultural and Biological Sciences, Environmental Sciences, and Social Sciences areas, and most of them were recently published (50 publications in 2018). In this journal, China is the country with the most papers on Climate Change, with 25 publications since 2006.

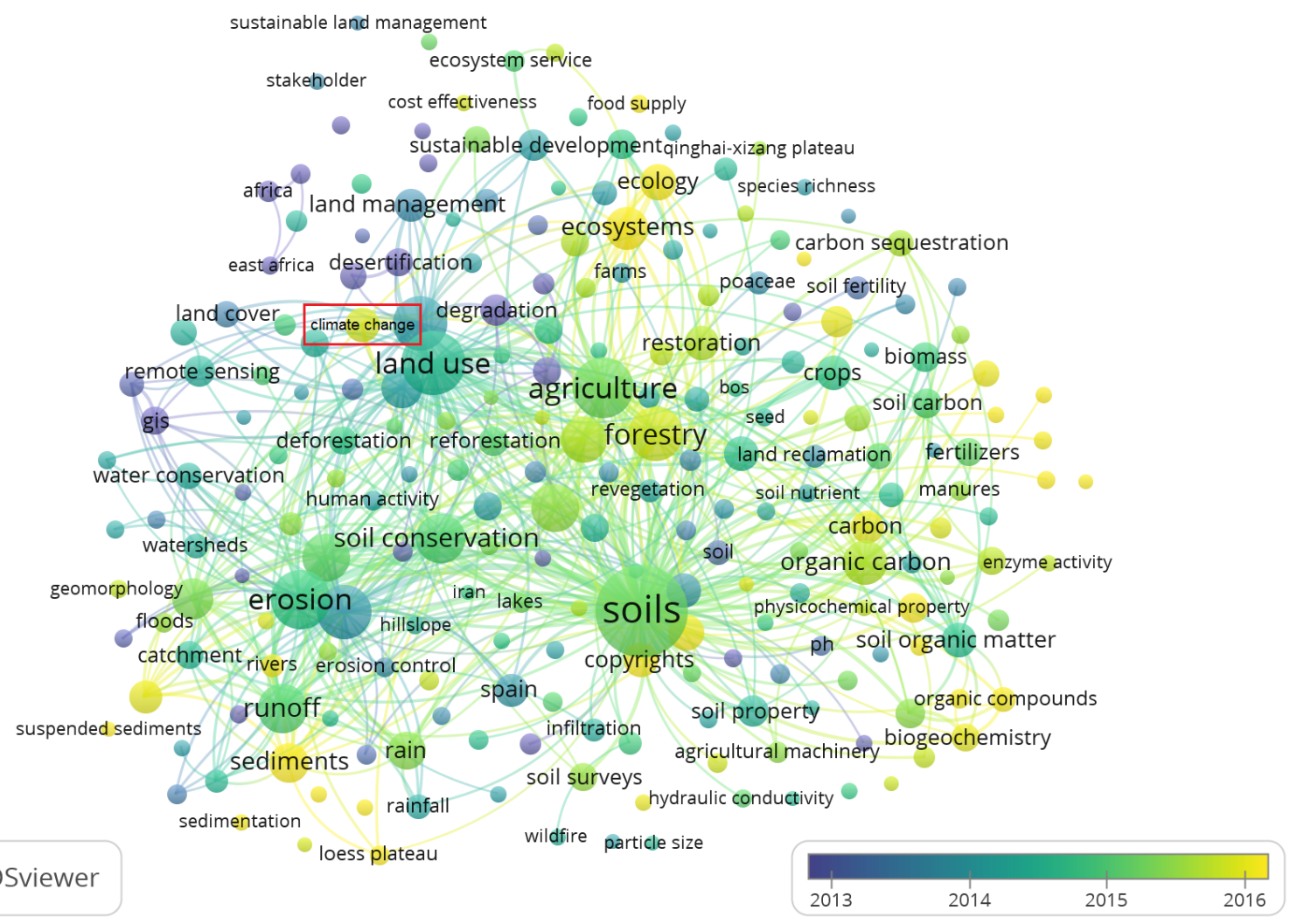

**Fig. 3:** Keywords from Land Degradation & Development journal papers from 2006 to 2018. Source: The authors, 2018.

III. Progress in Human Geography

Figure 4 shows the keyword from the *Progress in Human Geography* journal. Climate Change occurs 20 times and ranks 15th in the keyword occurrence list. The majority of those 20 papers were published in 2012 (6) and the countries with more publications are the United Kingdom (6), Australia (5), and the United States (4).

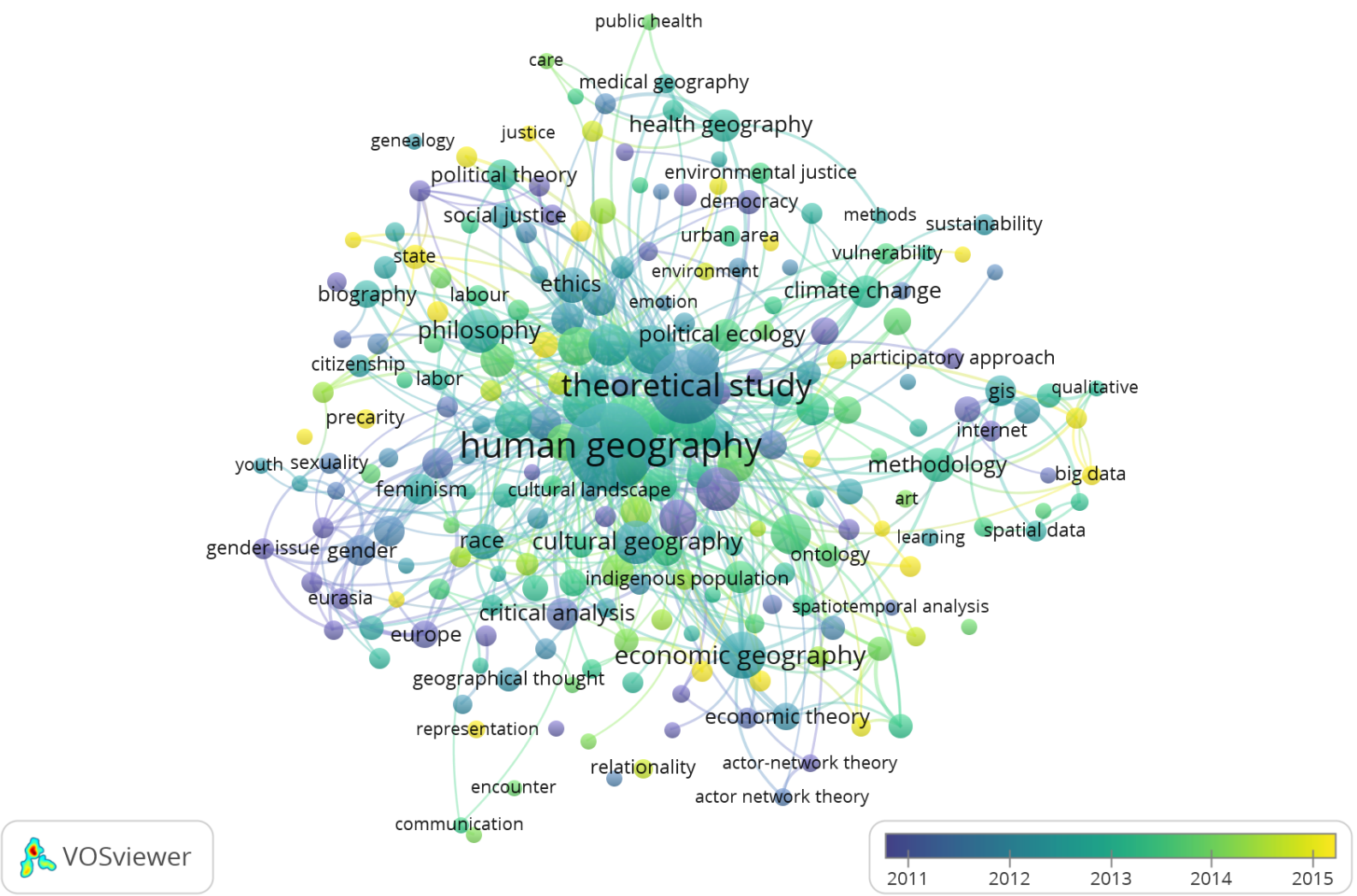

**Fig. 4:** Keywords from Progress in Human Geography journal papers from 2006 to 2018. Source: The authors, 2018.

## IV. Global Environmental Change – Human and Policy Dimensions

Another journal in which Climate Change is the most frequent keyword is *Global Environmental Change – Human and Policy Dimensions*, with 638 mentions. In 36 of these occurrences, publications are indexed only in Environmental Sciences. In the other 602 cases, they are indexed both in Environmental Sciences and Social Sciences areas. The majority of the 638 papers were published in 2014 (88) but, since 2006, they represent an expressive amount per year. In this journal, the United States (244), United Kingdom (194), and Australia (101) lead the rank of publications by country. Figure 5 shows the relation among all keywords indexed in the *Global Environmental Change – Human and Policy Dimensions* journal from 2006 to 2018.

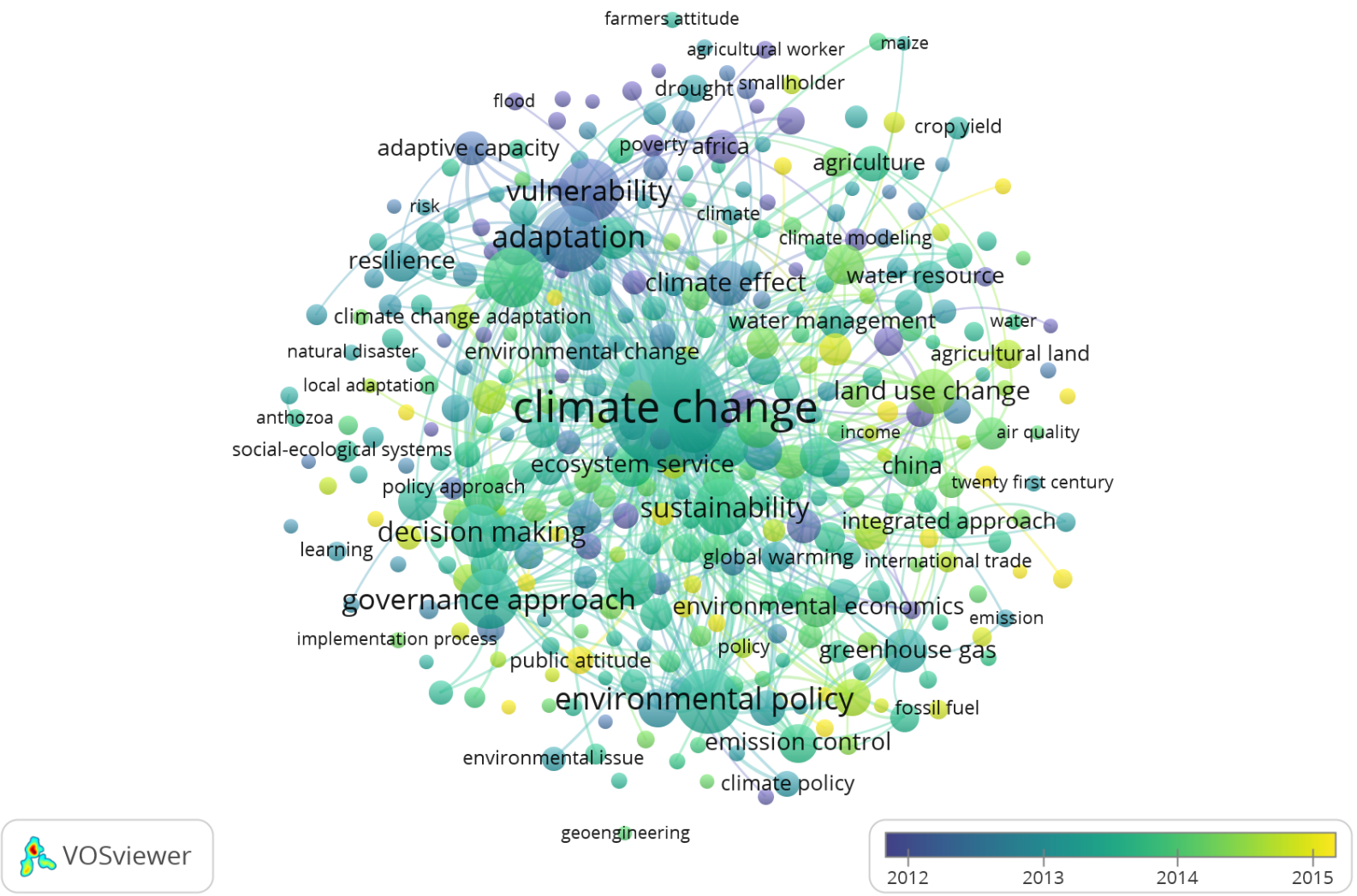

**Fig. 5:** Keywords from Global Environmental Change – Human and Policy Dimensions journal papers from 2006 to 2018. Source: The authors, 2018.

## V. Tourism Management

In the *Tourism Management* journal, the keyword Climate Change has 31 occurrences with 71 links to other 334 keywords as presented in the word cloud of Figure 6. All of these 31 papers are indexed in Business, Management and Accounting, and Social Sciences areas. They were produced mostly in Canada (9), Australia (7), and the United States (7).

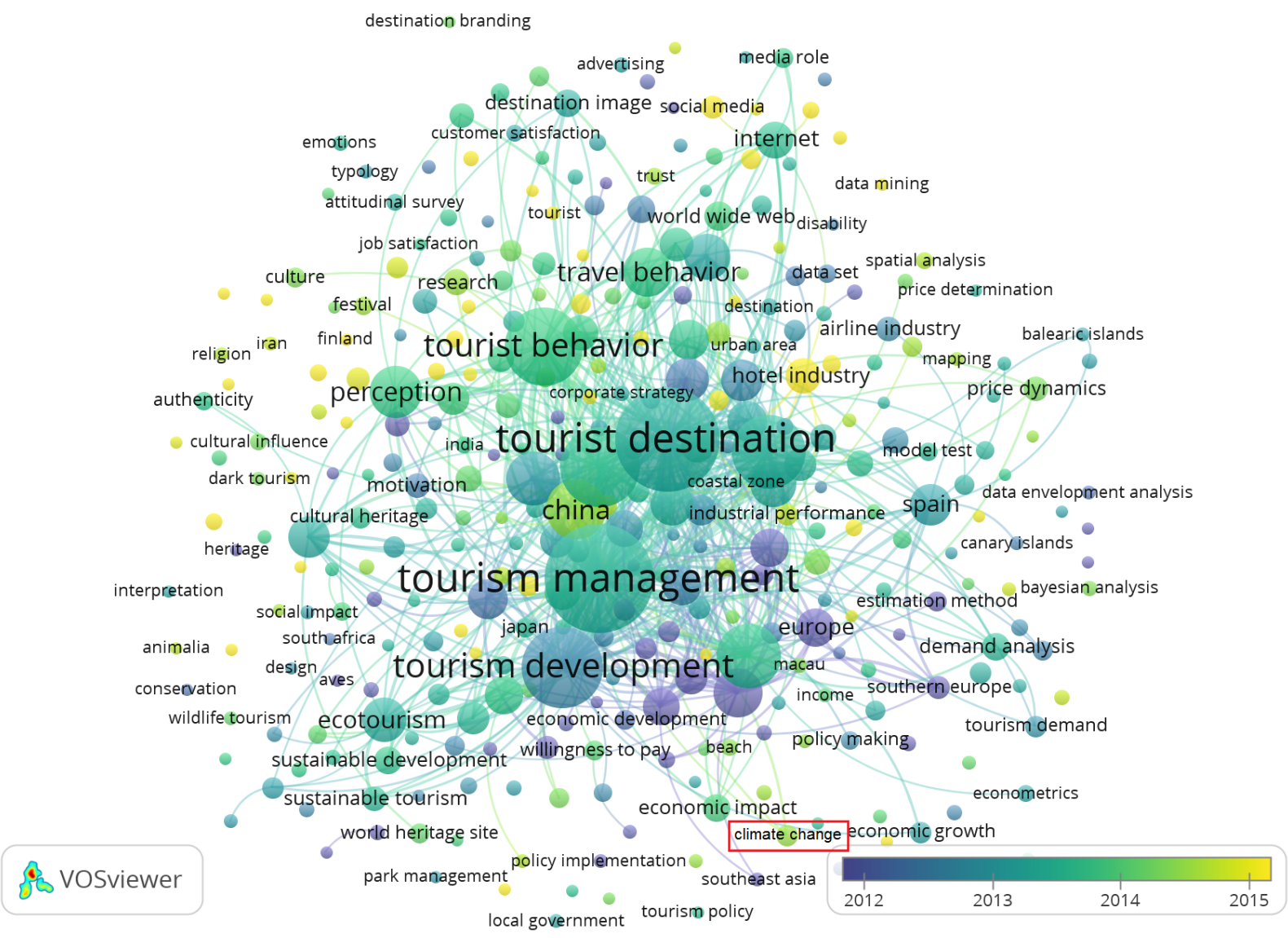

**Fig. 6:** Keywords from Tourism Management journal papers from 2006 to 2018. Source: The authors, 2018.

### 3.3 Climate Change production in the Social Sciences

The final step to understanding the production of knowledge about Climate Change and Social Sciences was to analyze all the 1,452 papers from 2007 to 2018 with the keyword Climate Change and indexed in Social Sciences[4] in the top-ranked journals. The 10th most cited keywords in these publications were: Climate Change (1,452 occurrences), Climate Effect (219); Environmental Policy (194); Adaptation (181); Vulnerability (171); Adaptive Management (170); United States (149); Greenhouse Gas (141); Anthropogenic Effect (123); Climate Modeling (112).

Until 2013, the publications followed a rising line, with a drop in 2014, a stable fluctuation thereafter, and peaking in 2016 with 192 publications, as Figure 7 shows. The ranking of publications by country (Figure 8) is led by the United States (617); followed by the United Kingdom (432); Australia (221); Germany (160); Netherlands (135); and Canada (117). The Global South has less representation as expected, with South Africa (36) in 16th, and Brazil occupying the 21st position with 22 publications.

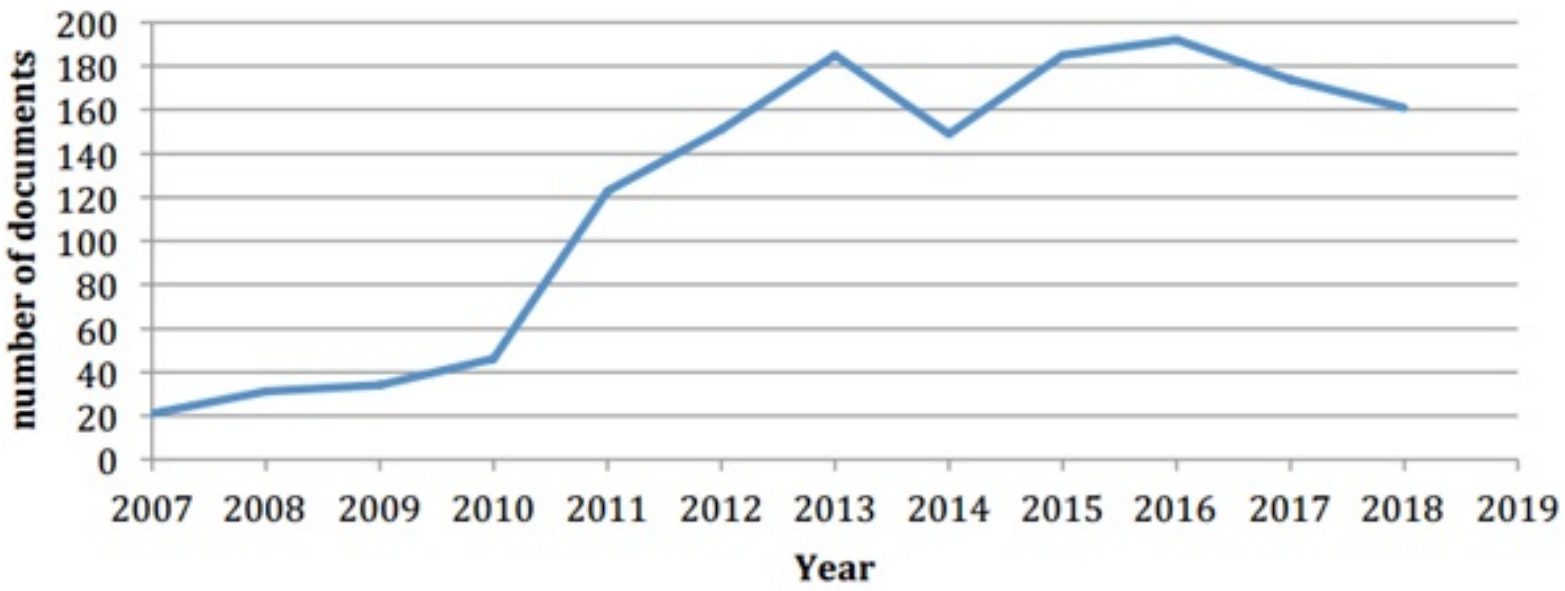

**Fig. 7:** Total number of publications per year. Source: The authors, 2018.

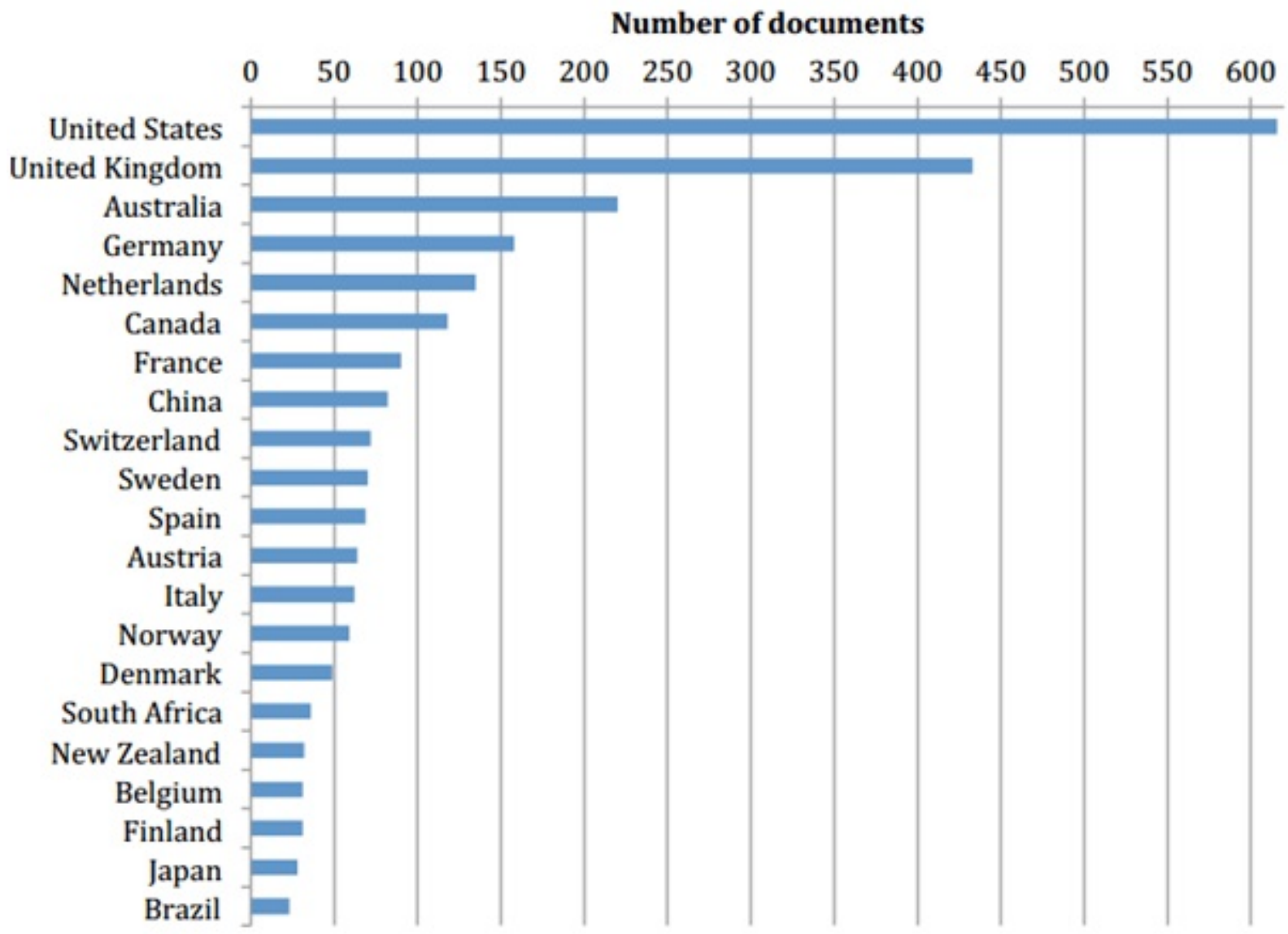

**Fig. 8:** Countries with the largest number of published documents. Source: The authors, 2018.

When we look to the top-cited authors in the selected papers (Figure 9), it is remarkable that none of them are strictly from the Social Sciences field or affiliated with a Social Sciences department, and they have, in general, a great number of works[5]. The best examples are the cases of professors Reto Knutti, Detlef van Vuuren, and Keywan Riahi, from the natural science/environmental science field. Professor Neil Adger, from the Geography Department, represents a rare case of a human-social science exponent at the Highly Cited Authors at the Clarivate Analytics Platform.

In terms of gender, there is a balance between the authors. Concerning their affiliations, the largest number of contributions came from University of East Anglia (67 papers); Wageningen University and Research Centre (56); University of Exeter (55); University of Oxford (47); and University of Leeds (38), which means that, from the top 5 universities that publish about Social Sciences and Climate Change, 4 are from the United Kingdom. This partly explains the position of the UK as the second country with more contributions (Figure 8).

Even though Brazil ranks only as of the twenty-first country on the list, it is remarkable that a Brazilian author, based at the University of Michigan, Maria Carmem Lemos, is on the ranking of authors with the largest number of published documents (Figure 9).

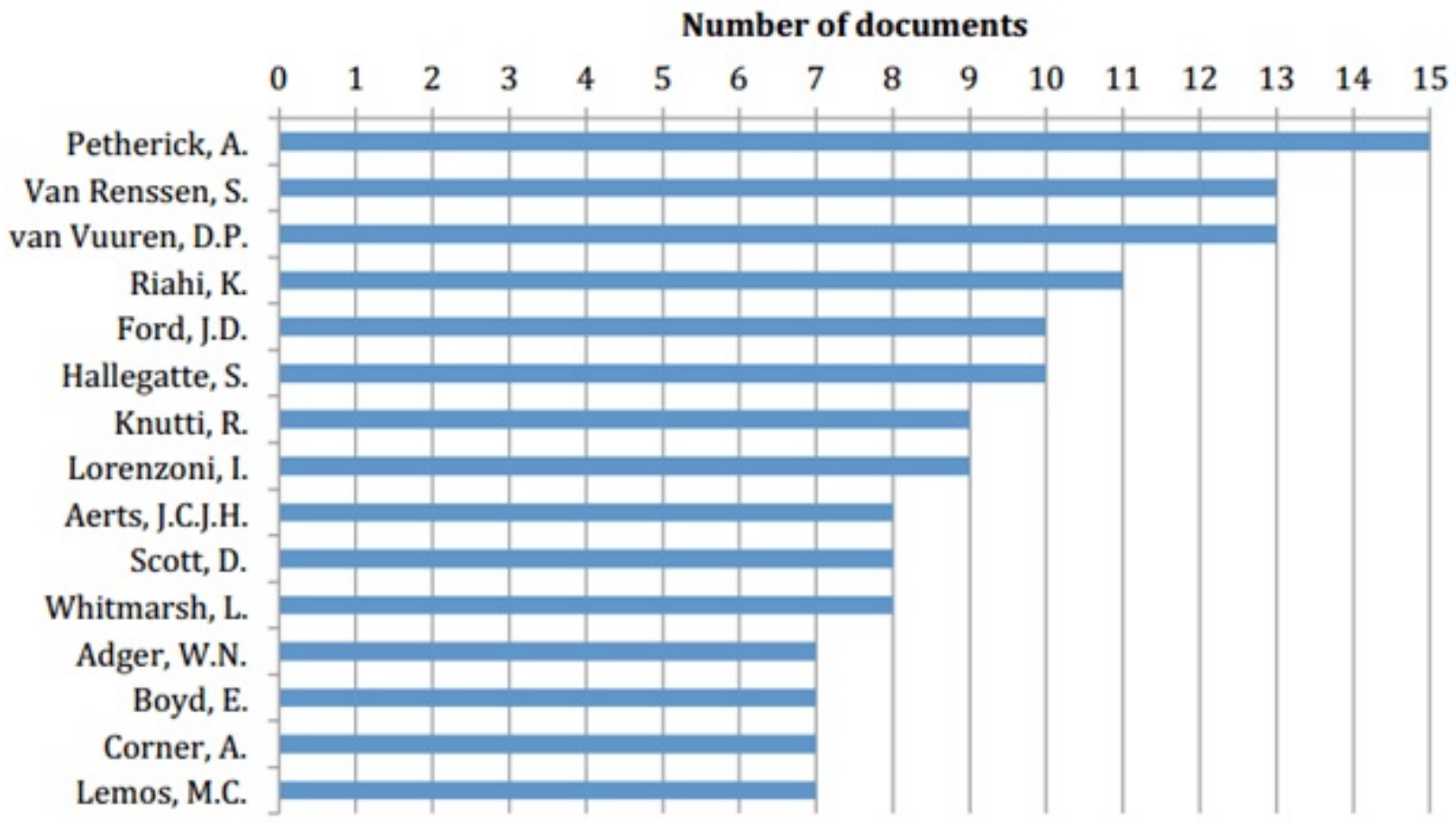

**Fig. 9:** Authors with the largest number of published documents. Source: The authors, 2018.

An important finding of our research is that the vast majority of articles about Climate Change indexed in Social Sciences are also indexed in Environmental Sciences (Figure 10).

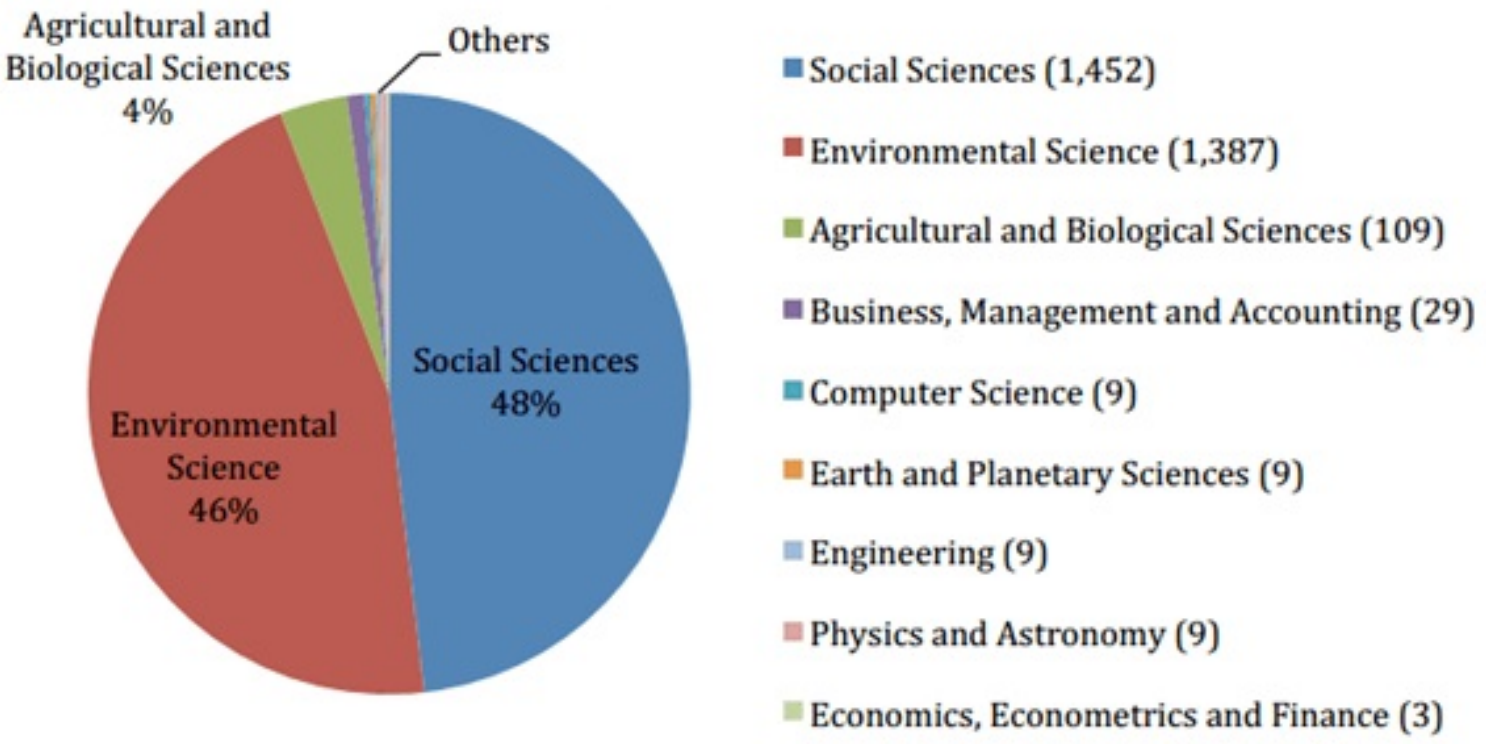

**Fig. 10:** Subject Area: Social Sciences and Environmental Science. Source: The authors, 2018.

## 4 Conclusion

The emergence of these new "hybrids understandings" (Jacobi, Rotondaro and Torres, 2019, our translation) in a more desirable interdisciplinary field could be the result of a continuous growth of an environmental dialogue (especially climate change), for example, inside the Social Sciences field. The difficulties to select only considerable/indexed information in Social Sciences are described in this research. Some journals, and also some academic researchers, are already more interdisciplinary in their current works.

The need of finding mechanisms that allow us to have a less fragmented view of reality imposes the challenge of strengthening interdisciplinary practices when dealing with problems such as the environmental "crisis" that is emerging on a planetary scale never seen before.

As a consequence, the challenge of breaking with the compartmentalization and diversity marginalization of knowledge is a relevant issue that involves a set of actors from the educational universe at all levels. It strengthens the possibility of fostering other rationalities for the engagement of different systems of knowledge, training, and professionalization. It consequently empowers content and knowledge based on values and practices indispensable to stimulate interest, engagement, and accountability (Jacobi, Giatti and Ambrizzi, 2015).

The production of knowledge must necessarily contemplate the interrelationships between natural and social environments, their subsistence, and conflicts with the dominant means of production. This includes the analysis of the process determinants, the role of the various actors involved, and the forms of social organization that increase the power of alternative actions. This is a perspective that strengthens a logic based on the transversality between knowledge, with an emphasis on socio-environmental sustainability, and management of common resources from a moral-ethical point of view.

For the actual construction of practices capable of structuring the foundations of sustainable societies, there is a need to strengthen communities of practice (Wenger, 1998) and Social Learning (Jacobi, 2011). These are as processes that allow increasing the number of people in the knowledge-building exercise and strengthening democratic communication channels to create and enhance interactions that bring substantial advances in the production of new social mobilization repertoires and practices for sustainability.

The recent merger of International Council for Science (ICSU) and International Social Science Council (ISSC) into International Council for Science (ICS) will be an important arena to observe the movements and pathways taken to the contribution – or not – of the Social Sciences field to the Climate Change debate, as well as the Future Earth Networks[6] and the Earth System Governance[7]. The contribution will likely grow in a more hybrid and interdisciplinary way, catalyzed by transnational networks more than by academic journals, which is very much desirable.

Finally, together with the growing presence of interdisciplinarity on the scientific production, arises the necessity of developing new methods and tools to understand and analyze how interdisciplinarity is present in a dialogue and how the knowledge of a specific area contributes to an interdisciplinary dialogue. This work demonstrates a way of evaluating the contribution from the Social Sciences field to the construction of new hybrids of understandings to the climate change dialogue through an innovative bibliometric research method.

**Acknowledgment**

We acknowledge financial support from Grants No. 2018/06685-9, 2019/05644-0, and 2019/18462-7 of the São Paulo Research Foundation (FAPESP). These grants are part of the FAPESP thematic project "Environment

Governance in São Paulo Macrometropolis in a climate variability context” No. 2015/03804-9.

---

**1** Available at https://council.science/current/press/international-science-council-holds-first-general-assembly-in-paris/. Accessed: 8 Mar. 2020.

**2** Available at https://jcr.clarivate.com/JCRLandingPageAction.action. Accessed: 8 Mar. 2020.

**3** See http://www.vosviewer.com/

**4** Some papers from Global Environmental Change-Human and Policy Dimensions journal and ISPRS Journal of Photogrammetry and Remote Sensing are not indexed in the Social Sciences subject area and were not considered.

**5** Google Scholar database.

**6** See https://network.futureearth.org/home

**7** See https://www.earthsystemgovernance.org/